\documentclass[prl,twocolumn,aps,superscriptaddress]{revtex4-1}
\usepackage{graphicx}% Include figure files
\usepackage{dcolumn}% Align table columns on decimal point
\usepackage{bm}% bold math
\usepackage{color}
\usepackage{blindtext}
\usepackage[utf8]{inputenc}
\usepackage{appendix}
\usepackage{lipsum}
\usepackage{xr}
\usepackage{xcolor}
\usepackage{siunitx}
\usepackage{multirow}

\usepackage[colorlinks]{hyperref}% add hypertext capabilities
\hypersetup{%
	plainpages=true,
	breaklinks=true,% not default in dvips mode, so we must specify
	hypertexnames=false,%not ideal, but needed when pagenums duplicate (`i' vs. `1')
	pageanchor=true,
	colorlinks=true,
	linkcolor={blue},
	citecolor={blue},
	urlcolor={blue},
	anchorcolor={black}
}
\begin{document}
	\title{Single-mode magnon-polariton lasing and amplification controlled by dissipative coupling}
	\author{Zi-Qi Wang}
	\affiliation{Zhejiang Key Laboratory of Micro-Nano Quantum Chips and Quantum Control, School of Physics, and State Key Laboratory for Extreme Photonics and Instrumentation, Zhejiang University, Hangzhou 310027, China}
	\author{Zi-Yuan Wang}
	\affiliation{Zhejiang Key Laboratory of Micro-Nano Quantum Chips and Quantum Control, School of Physics, and State Key Laboratory for Extreme Photonics and Instrumentation, Zhejiang University, Hangzhou 310027, China}
	\author{Yi-Pu Wang}
	\email{yipuwang@zju.edu.cn}
	\affiliation{Zhejiang Key Laboratory of Micro-Nano Quantum Chips and Quantum Control, School of Physics, and State Key Laboratory for Extreme Photonics and Instrumentation, Zhejiang University, Hangzhou 310027, China}
	\author{J. Q. You}
	\email{jqyou@zju.edu.cn}
	\affiliation{Zhejiang Key Laboratory of Micro-Nano Quantum Chips and Quantum Control, School of Physics, and State Key Laboratory for Extreme Photonics and Instrumentation, Zhejiang University, Hangzhou 310027, China}
	\affiliation{College of Optical Science and Engineering, Zhejiang University, Hangzhou 310027, China}

\begin{abstract}
We demonstrate single-mode lasing of magnon polaritons in a cavity magnonic system enabled by dissipative coupling between two passive modes, microwave cavity mode and magnon mode in a ferrimagnetic spin ensemble. The cavity mode is partially compensated through a feedback circuit, which reduces its linewidth but retains its dissipative nature. By tuning the compensation strength and dissipative coupling strength, we reach a system cooperativity of unity, marking the lasing threshold and the formation of a zero-linewidth polariton mode. This mode also corresponds to a perfect Friedrich–Wintgen bound state in the continuum. Further increase of the cooperativity drives the system into the strong dissipative coupling regime, where magnon-polariton amplification arises between two real-frequency scattering poles. These results reveal that dissipative coupling cooperativity carries a clear physical meaning and serves as a key parameter for controlling phase transitions. Dissipative coupling offers an alternative paradigm for tailoring light–matter interactions, paving the way for advances in both information processing and quantum technologies.
\end{abstract}

\maketitle
\textit{Introduction}.---Dissipative coupling has garnered considerable interest owing to its capacity to facilitate unconventional phenomena in light–matter interactions~\cite{Kyriienko-14,Clerk-15,Xiao-16,Bernier-18,Ding-19,Zou-22,jiteng-22}. A typical dissipatively coupled two-mode system is characterized by level attraction in dispersion and repulsion in linewidth~\cite{Harder-18,Bernier-18,Ying-20,Grigoryan-19,Nair-21,guo-23,Carrara-24,Li-24,VanLoo-13}, where one mode experiences increased dissipation while the other undergoes a reduction. The latter arises from destructive interference between dissipation channels. This form of collective dissipation underpins a variety of dissipation-suppressed phenomena, including dark states~\cite{Dicke-54,Zanner-22,Tiranov-23}, decoherence-free subspaces~\cite{Lidar-98,Duan-97,Kwiat-00}, and the Friedrich--Wintgen bound state in the continuum (FW-BIC)~\cite{Hsu-2016,FWBIC,Wanghu-20}. Among them, the BIC exemplifies how unavoidable intrinsic losses constrain experimental implementations to quasi-BICs with finite damping~\cite{Kang-23,Yang-23}. By actively compensating for intrinsic dissipation, one can achieve complete linewidth cancellation, thereby realizing a \textit{perfect} FW-BIC. This zero-damping mode simultaneously marks the threshold for dissipative-coupling-induced lasing. Unlike coherent-coupling-based lasers~\cite{Feng-2014,Hodaei-2014}, dissipative-coupling approach requires neither a net gain mode nor rigorous gain–loss balance, and enables stable single-mode operation that can be conveniently controlled via the dissipative coupling strength and mode detuning. From the coupled-mode theory perspective, the lasing threshold corresponds precisely to the condition where the cooperativity reaches unity, $C_\Gamma = \Gamma^2 / (\gamma_{\rm c} \gamma_{\rm m}) = 1$, with $\Gamma$ denoting the dissipative coupling strength and $\gamma_{\rm c,m}$ the total dissipation rates of the two modes. However, as dissipative coupling is mediated through loss processes, the associated coupling strength is fundamentally limited, making it difficult to reach or surpass unity cooperativity~\cite{Clerk-15,Yao-19,Wanghu-20,Weichao-19,Xia-18}. To date, dissipative coupling-induced lasing has only been validated in the phonon platform~\cite{jiteng-22}. Moreover, while strong coupling in coherent systems has been extensively studied, the \textit{strong dissipative} coupling ($\Gamma>\gamma_{\rm c,m}$) requires further investigation to drive a paradigm shift in light--matter interactions. 

The emerging field of cavity magnonics~\cite{Rameshti-22,Yuan-22,Zhedong-19,Boventer-18} and waveguide magnonics~\cite{Ziqi-22,Jinwei-20,Xiong-23}, based on coupled magnons and microwave photons, has attracted increasing interest in recent years~\cite{Huebl-13,Xufeng-14,Tabuchi-14,Tobar-14,Bhoi-14,Bai-15,Tabuchi-15,Dengke-15,Xufeng-15,Tabuchi-15,Dengke-17,Wang-18,WangB-16,Wang-2019,Yao-2023,Jinwei-23,Xufeng-24,Jinwei-2024,YiLi-2019,Wang-2023,Jin-2023,Zhang-2021,Luo-2021}. The tunable frequency~\cite{note1} and dissipation~\cite{Ying-20,Yao-21JAP,Yao-19CP,Tao-24} of magnon modes endow this system with distinct advantages for exploring dissipation-enabled non-Hermitian physics~\cite{Dengke-17,Ying-20,Harder-17,Harder-18,Guoqiang-19,Peng-19,Tao-20,Jing-21,Jinwei-2024,Qian-24}. Particularly, the gyromagnetic nature of magnon modes offers exceptional opportunities for studying topological photonics~\cite{Chong-09,Liu-22,Chen-24}. By further coupling magnons with optical photons~\cite{Tang-16,Nakamura-16,Nakamura-16-2,Haigh-16,weijiang24}, phonons~\cite{Tang-16-2,Li-18,Potts-21,Shen-22,Dong-23}, and superconducting qubits~\cite{Tabuchi-15,Dany-17,Dany-20,Wolski-20,Xu-23,Xu-24,Yan-24}, a platform of magnon-based hybrid quantum systems is gradually taking a shape~\cite{Nakamura-19APE}. These advancements have been primarily built upon \textit{coherently coupled} magnonic systems. In parallel, dissipative coupling between magnons and microwave photons has also been experimentally demonstrated~\cite{Harder-18,Bhoi-19,Xia-18,Wang-2019,Ying-19,Pengchao-19,Yuan-20,Wanghu-20}, but most previous studies have remained in the weak coupling regime.

	\begin{figure}[ht]
	\includegraphics[width=0.47\textwidth]{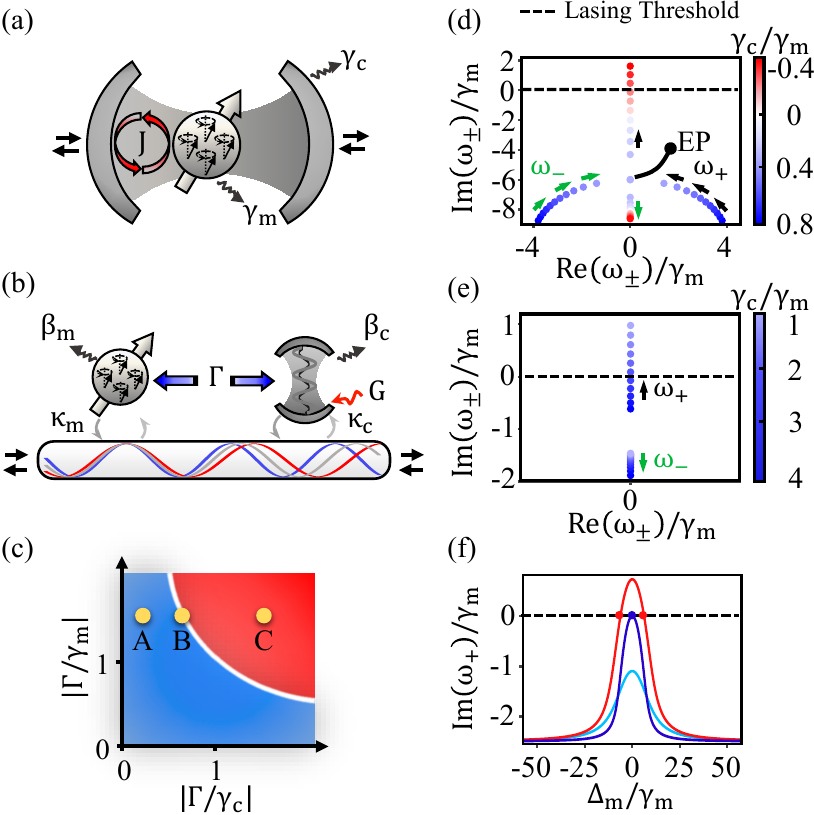}
	\caption {(a),(b) Schematic diagrams of the coherently (a) and dissipatively (b) coupled cavity magnonic systems.
		(c) Diagram of the dissipative coupling regimes. The white line corresponds to the unity cooperativity, which divides the areas into passive (blue) and active (red) regimes. The points A, B, C represent three typical conditions studied in experiments.
		(d),(e) Evolution of eigenfrequencies in the complex plane for the coherent coupling system (d) and dissipative coupling system (e) as the decay rate \(\gamma_{\rm{c}}\) is varied.
		(f) Imaginary part of $\omega_{+}$ versus cavity photon-magnon detuning $\Delta_{\rm{m}}$ in different coupling regimes. The cyan, blue, and red curves correspond to points A, B and C in (c), respectively.
	}
	\label{fig1}
\end{figure}	

In this work, we overcome this limitation by developing an active cavity magnonic system, enabling the cooperativity of the dissipatively coupled magnon-photon system to gradually exceed unity, ultimately reaching the strong dissipative coupling regime. Under the condition of unity cooperativity, we clearly observe single-mode magnon-polariton lasing, and also the emergence of a perfect FW-BIC. Its ultra-narrow linewidth is attractive for high-\textit{Q} applications, such as precision sensing, narrowband filters, and quantum transducers. As the system transitions into the strong dissipative coupling regime, magnon-polariton lasing occurs at specific positive or negative detunings between the magnon mode and the cavity mode. Within the frequency range between these detunings, we observe amplification and instability of the magnon polariton. Furthermore, we demonstrate that the lasing threshold can be effectively reduced by increasing the dissipative coupling strength. These findings establish the dissipative cooperativity as a key control parameter, enabling lasing, amplification and spectral shaping without fine-tuned gain-loss balance conditions.

\textit{System and concepts.}---To determine the lasing condition from the scattering perspective, we consider the scattering pole, a complex-frequency singularity of the system's scattering matrix~\cite{Feng-2014,Hodaei-2014,Chong-10,Wang-21}. The pole indicates a divergent response of the scattering and inherently corresponds to the eigenvalue of the system’s effective non-Hermitian Hamiltonian. The lasing physically occurs when such a pole moves onto the real axis, where the output intensity diverges at a real driving frequency. At this point, at least one eigenmode has a real eigenfrequency. For coherently and dissipatively coupled cavity magnonic systems, schematic illustrations are shown in Figs.~\ref{fig1}(a) and \ref{fig1}(b), respectively. The coupled system comprises cavity and magnon modes with frequencies $\omega_{\rm{c,m}}$, where the subscripts $`{\rm c}$' and $`{\rm m}$' denote the cavity mode and magnon mode, respectively. Their total dissipation rates $\gamma_{\rm{c,m}}$ consist of the radiative and intrinsic parts, $\kappa_{\rm{c,m}}$ and $\beta_{\rm{c,m}}$. In both cases, the interaction can lead to the formation of two polariton modes with complex eigenfrequencies,
	\begin{equation} \label{polartion}
		\omega_{\pm}=\frac{1}{2}\left[\omega_{\rm{c}}+\omega_{\rm{m}}-i\left(\gamma_{\rm{c}}+\gamma_{\rm{m}} \right) \pm i\delta\right],
	\end{equation}
where $\delta\equiv\delta_{\rm{J,\Gamma}}=\sqrt{\mp4C_{\rm{J,\Gamma}}\gamma_{\rm{c}}\gamma_{\rm{m}}-\left[i\left(\gamma_{\rm{c}}-\gamma_{\rm{m}}\right)-\Delta_{\rm{m}}  \right]^{2}}$. In the coherent (dissipative) coupling case, $C_{J(\Gamma)}=\frac{J(\Gamma)^2}{\gamma_{\rm{c}}\gamma_{\rm{m}}}$ is the system cooperativity, $J~(\Gamma=\sqrt{\kappa_{\rm{c}}\kappa_{\rm{m}}})$ is the coherent (dissipative) coupling strength \cite{SM}, and $\Delta_{\rm{m}}=\omega_{\rm{m}}-\omega_{\rm{c}}$ is the frequency detuning between the two modes. 

With fixed coupling strengths, $\Delta_{\rm{m}} = 0$, and $\gamma_{\rm{m}} > 0$, we examine how the eigenfrequencies evolve as $\gamma_{\rm{c}}$ decreases, as illustrated in Figs.~\ref{fig1}(d) and \ref{fig1}(e). In the well-studied coherent case, the polariton begins lasing [$\rm{Im}(\omega_{+}) = 0$] only when cavity mode possesses a net gain, i.e., \(\gamma_{\rm{c}} < 0\). However, in a dissipative coupling system, one polariton mode can achieve lasing even with the cavity mode remaining lossy but compensated. This indicates that dissipative coupling can induce lasing without the need of pre-established cavity gain. Using Eq.~(\ref{polartion}), we find that a zero-linewidth polariton mode is readily obtained with $C_{\Gamma}=1$, which serves as the threshold for dissipative-coupling-induced polariton lasing and perfect FW-BIC. Accordingly, we interpret the dissipative coupling regimes shown in Fig.~\ref{fig1}(c). The white line indicating $C_{\Gamma}=1$ divides the areas into the passive regime (blue region) and the amplifying regime (red region). In dissipatively coupled systems, the intrinsic damping of the constituent modes typically prevents the cooperativity from approaching unity, thereby limiting previous studies to the blue region. We plot $\rm{Im}(\omega_{+})$ versus mode detuning for three different values of cooperativity in Fig.~\ref{fig1}(f), corresponding to points A, B and C in Fig.~\ref{fig1}(c), respectively. It can be seen that a single real-valued pole appears at the magnon-photon resonance when $C_{\Gamma}=1$. As $C_{\Gamma}$ increases further, the single pole splits into a pair of real-valued poles at opposite magnon-photon detunings, with polariton amplification occurring between them. Since only one scattering pole intersects the real axis at a specific frequency, the system inherently favors stable single-mode operation.
	\begin{figure}
	\includegraphics[width=0.46\textwidth]{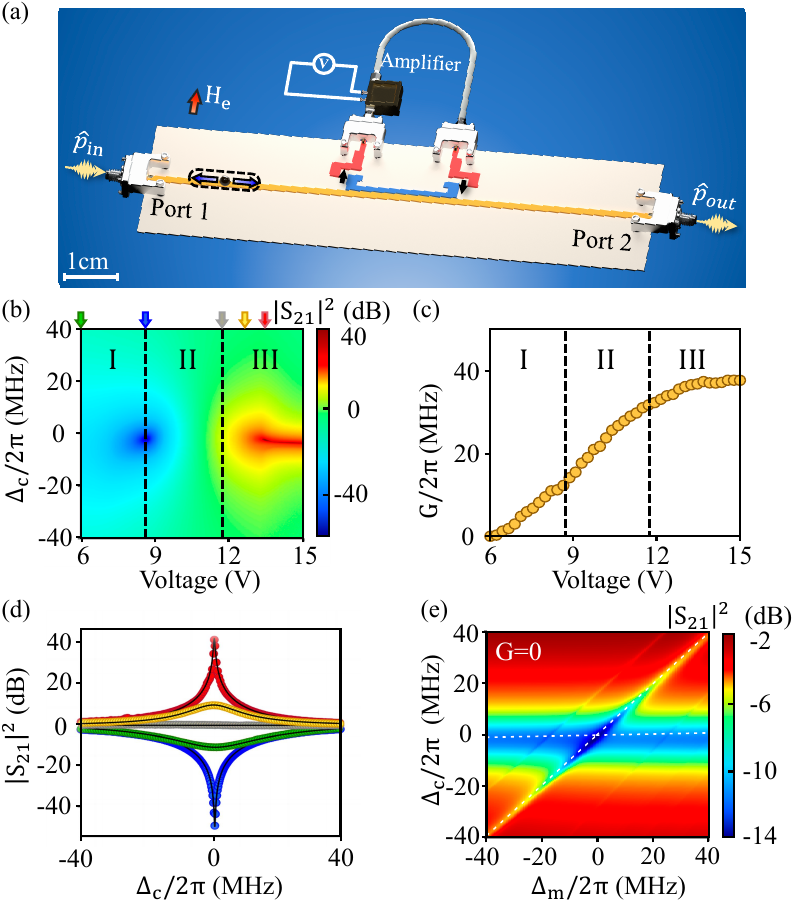}
	\caption {(a) Schematic of the experimental device, where the split-ring cavity (blue) is compensated by an active circuit (red) and side-coupled with the waveguide (yellow). The YIG sphere (black) is moved along the waveguide. (b) Measured device transmission versus bias voltage $V$ without the YIG sphere .
		(c) The compensation rate $G$ fitted from the transmission mapping shown in (b).
		(d) Typical transmission spectra extracted from (b). The green, blue, gray, yellow, and red curves correspond to $G = 0$, $\beta_{\rm{c}}$, $\kappa_{\rm{c}}/2 + \beta_{\rm{c}}$, $2\kappa_{\rm{c}}/3 + \beta_{\rm{c}}$, and $\kappa_{\rm{c}} + \beta_{\rm{c}}$, respectively.
		(e) Measured transmission of the dissipatively coupled cavity magnonic system as a function of the bias magnetic field without compensation ($G = 0$). The white dashed lines denote the uncoupled cavity and magnon modes.
	}
	\label{fig2}
\end{figure}	

\textit{Experimental results.}---To achieve unity cooperativity and strong dissipative coupling, we first construct a microwave feedback loop to compensate the cavity dissipation. Our device consists of a half-wavelength cavity (distorted split-ring resonator, blue) and a feedback circuit (red) connected to an amplifier [Fig.~\ref{fig2}(a)]. To effectively reduce the cavity decay rate, the cavity and feedback circuit are designed to ensure the feedback photons remain in phase with the cavity photons. The microstrip waveguide (yellow) is side-coupled with the ring resonator, and separated from the active circuit. For this feedback-cavity-waveguide system, the transmission coefficient can be expressed as \cite{SM}
	\begin{equation} \label{s21_single}
		S_{21}(\Delta_{\rm{c}})=\frac{i\Delta_{\rm{c}}-\beta_{\rm{c}}+G} {i\Delta_{\rm{c}}-\gamma_{\rm{c}}+G},
	\end{equation}
where $\Delta_{\rm{c}}=\omega-\omega_{\rm{c}}$ is frequency detuning between the probe field and cavity mode. The compensation rate to the cavity mode $G$ can be continuously tuned by adjusting the bias voltage $V$ applied to the amplifier. Figure~\ref{fig2}(b) shows the transmission spectra measured versus $G$ at room temperature, using a $-50$~dBm input signal. By fitting the transmission spectra with Eq.~(\ref{s21_single}), the compensation rate $G$ versus voltage is plotted in Fig.~\ref{fig2}(c), showing a monotonic increase. The radiative damping and frequency of the cavity mode remain almost unchanged during this adjustment~\cite{SM}. $\beta_{\rm{c}}/2\pi$ and $\kappa_{\rm{c}}/2\pi$ of the \textit{bare} cavity mode are fitted to be 12.2 MHz and 28.6 MHz, respectively, corresponds to the green curve in Fig.~\ref{fig2}(d). 

Based on how the compensation rate $G$ relates to $\beta_{\rm c}$ and $\kappa_{\rm c}$, we divide the mapping in Fig.~\ref{fig2}(b) into three regions. In region I $(0 \le G \le \beta_{\rm{c}})$, the transmission valley becomes deeper with increasing voltage, which will decrease to zero when the compensation rate equals to the intrinsic damping ($G = \beta_{\rm{c}}$), as shown by the blue curve in Fig.~\ref{fig2}(d). As the bias voltage increases, the system enters region II ($\beta_{\rm c} < G \leq \kappa_{\rm c}/2 + \beta_{\rm c}$), where $\left|S_{21}(\omega_{\rm c})\right|$ increases with $G$ and reaches unity at $G = \kappa_{\rm c}/2 + \beta_{\rm c}$, causing the transmission dip to gradually diminish and eventually flatten out, as shown by the gray line in Fig.~\ref{fig2}(d). The peculiar physics we aim to demonstrate occurs in region III ($\kappa_{\rm{c}}/2+ \beta_{\rm{c}} < G \le \gamma_{\rm{c}}$), where $|S_{21}(\omega_{\rm{c}})|$ exceeds unity, allowing the amplification around the cavity mode. The lineshape transforms into a peak [yellow curve in Fig.~\ref{fig2}(d)], and grows higher with further increasing the voltage. Eventually, the peak stabilizes at a high intensity (40 dB) regardless of the additional voltage. For the red curve with a pronounced narrow peak shown in Fig.~\ref{fig2}(d), the cavity dissipation is reduced to zero ($G = \gamma_{\rm{c}}$), causing $|S_{21}(\omega_{\rm{c}})|$ to diverge, corresponding to a scattering pole. The minimum voltage satisfying this condition marks the lasing threshold, which is 12.44 V. The unchanged lineshape at higher voltages arises from adaptive compensation of cavity loss by the feedback circuit, resembling self-sustained behavior reported previously~\cite{Yao-2023}. At this limit ($G = \gamma_{\rm c}$), the cavity mode becomes lossless but does not exhibit net gain. Consequently, when coherently coupled to a lossy magnon mode, magnon-polariton lasing still cannot occur.

\begin{figure}
	\includegraphics[width=0.46\textwidth]{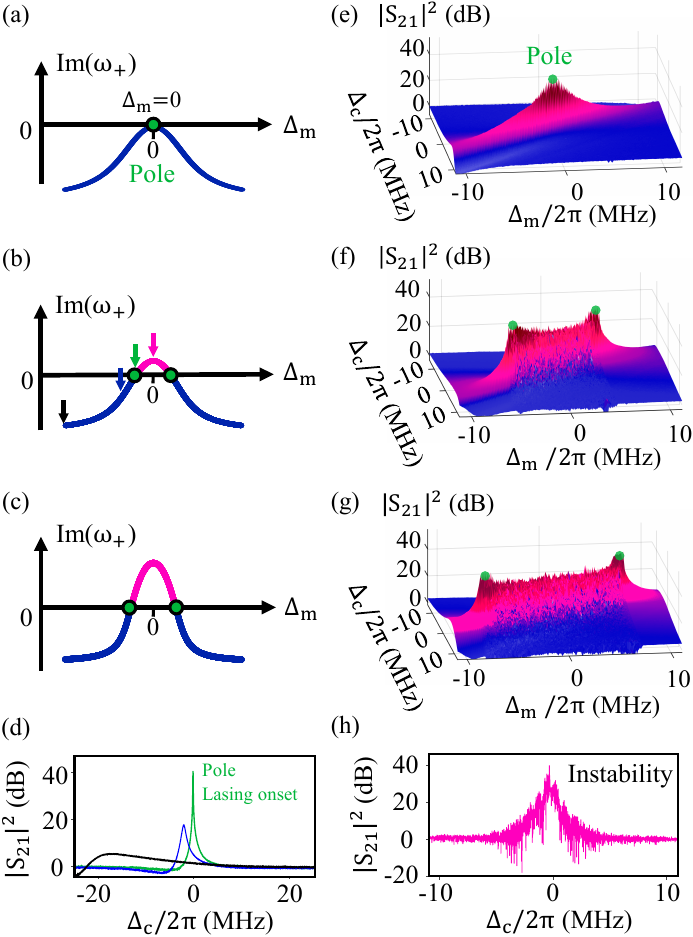}
	\caption {(a)-(c) Calculated $\rm{Im}(\omega_{+})$ versus mode detuning $\Delta_{\rm{m}}$ for different compensation rates $G$. The blue (pink) segment indicates the $\rm{Im}(\omega_{+})$ greater (less) than zero. The green dots represent the real-valued poles (lasing threshold $\rm{Im}(\omega_{+})=0$). The parameters used in these calculations match the experimental results shown in (e) to (g).		
		(e)-(g) Experimentally measured transmission spectra plotted as 3D diagrams versus mode detuning $\Delta_{\rm{m}}$ and cavity field detuning $\Delta_{\rm{c}}$.
		(d),(h) Spectra extracted from (f) at the positions marked by the arrows in (b). The black, blue, and ultrasharp green curves in (d) correspond to conditions far below, near, and precisely at the lasing threshold, respectively. The chaotic lineshape (pink) in (h) corresponds to the magnon-polariton amplification and instability.
	}
	\label{fig3}
\end{figure}

Next, we realize the dissipative coupling between the cavity mode and magnon mode. The magnon mode is supported by a 1 mm-diameter single-crystal yttrium iron garnet (YIG) sphere, which is commercially available~\cite{http://www.ferrisphere.com}. When the sphere is magnetized, the magnon frequency can be linearly adjusted by the bias magnetic field. The waveguide-mediated interaction between the cavity mode and magnon mode is determined by the cumulative propagating phase $\varphi$ between them~\cite{VanLoo-13,Ziqi-22}, which can be accurately adjusted by changing the relative distance between the YIG sphere and cavity. The sphere is glued to a cantilever mounted on a stepper motor, allowing precise spatial positioning along the waveguide. When the propagating phase is an integer multiple of $\pi$ ($J = 0$), purely dissipative coupling is established. With the compensation $G$, the system cooperativity $C_{\rm{\Gamma}}$ is modified as $\frac{\Gamma^2}{\gamma_{\rm{e}}\gamma_{\rm{m}}}$, where $\gamma_{\rm{e}} = \gamma_{\rm{c}} - G$. Thus, we can readily adjust $ C_{\rm{\Gamma}} $ to reach unity or even greater.

We first measure the transmission versus mode detuning $\Delta_{\rm{m}}$ without compensation ($G = 0$), as shown in Fig.~\ref{fig2}(e). The spectrum does not exhibit obvious level attraction. The fitted values of $\kappa_{\rm{m}}/2\pi$ and $\beta_{\rm{m}}/2\pi$ are 1.31~MHz and 1.53~MHz, respectively. Clear level attraction appears after the dissipative coupling strength is enhanced~\cite{SM}. It is worth noting that level attraction can also arise from the interference between two coherent driving tones~\cite{Boventer-2019,Boventer-2020,Gardin-2025,Xia-18}. However, this mechanism is fundamentally different from the traveling-wave-mediated dissipative coupling employed in our work. The polariton lasing cannot occur as the system is currently at point A in Fig.~\ref{fig1}(c). Then, we set the compensation rate to $G/2\pi =$ 27.6 MHz, corresponding to a bias voltage of 7.86 V. At this stage, system cooperativity approaches unity [point B in Fig.~\ref{fig1}(c)], where the effective dissipation rate of the cavity mode $\gamma_{\rm{e}}/2\pi$ is 13.2 MHz. The measured transmission spectra at varying bias magnetic fields are presented in a 3D diagram in Fig.~\ref{fig3}(e). In the large detuning condition, both the cavity mode and magnon mode function as \textit{lossy modes}. When the magnon mode is tuned to resonate with the cavity mode, the coupled system satisfies the lasing threshold and perfect FW-BIC condition [green point in Fig.~\ref{fig3}(a)]. At this point, the imaginary part of the corresponding eigenstate $\text{Im}(\omega_{+})$ is zero, while $\text{Im}(\omega_{-})$ is far from zero, ensuring the stable single-mode lasing operation.

	\begin{figure}
	\includegraphics[width=0.41\textwidth]{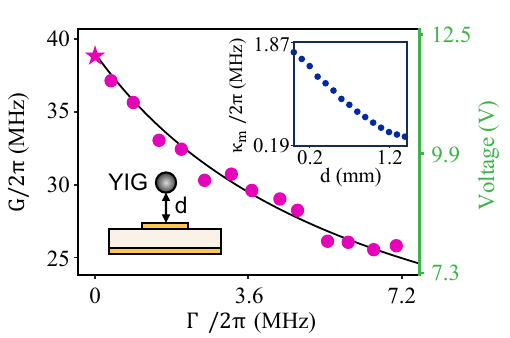}
	\caption {Threshold of compensation rate and bias voltage for generating magnon-polariton lasing at different dissipative coupling strengths $\Gamma$. The star indicates the cavity compensated to lasing without coupling to the magnon mode. The solid curve is the theoretical fit. The insets show the variation of $\kappa_{\rm{m}}$ as a function of the distance $d$ between the YIG sphere and the waveguide plane.
	}
	\label{fig4}
\end{figure}

With increasing compensation, $\text{Im}(\omega_{+})$ exceeds zero within a certain detuning range, as depicted by the pink curve in Fig.~\ref{fig3}(b). In this range, $\text{Im}(\omega_{+}) > 0$, magnon-polariton amplification is achieved, as shown by the chaotic spectra range between two poles in Fig.~\ref{fig3}(f). The transmission spectrum at one of the poles is plotted as the green curve in Fig.~\ref{fig3}(d), and the corresponding characterization of the perfect FW-BIC is provided in the supplementary materials~\cite{SM}. The non-Lorentzian lineshape indicates that the lasing occurs at non-resonant positions ($\Delta_{\rm{m}} \neq 0$). Compared to the spectra at far and near threshold detunings (black and blue curves) in Fig.~\ref{fig3}(d), the distinctly narrowed linewidth of the green curve provides clear evidence of lasing. The transmission spectrum of the magnon-polariton amplification is shown in Fig.~\ref{fig3}(h). The chaotic lineshape suggests that instability is stimulated. By further increasing the bias voltage, the reduced effective decay rate of the cavity ($\gamma_{\rm{e}}/2\pi = 4$ MHz) allows the system to enter the strong dissipative coupling regime [$\Gamma > \{\gamma_{\rm{m}}, \gamma_{\rm{e}}\}$]. Two transmission poles spread further apart [Fig.~\ref{fig3}(c)], broadening the frequency range of magnon-polariton amplification. The experimental results are depicted in Fig.~\ref{fig3}(g), showing that the distance between the two poles becomes approximately 17 MHz. 

To further illustrate the influence of dissipative coupling on the lasing threshold, we vary the dissipative coupling strength by changing radiative damping rate $\kappa_{\rm{m}}$ of the magnon mode. This is achieved by adjusting the distance of the YIG sphere from the device plane [inset of Fig.~\ref{fig4}]. As shown in Fig.~\ref{fig4}, we plot the compensation rate that enables the cooperativity to reach unity (lasing threshold) versus the dissipative coupling strength $\Gamma$. The corresponding bias voltage is on the right axis. It can be seen that dissipative coupling effectively reduces the lasing threshold and the experimental results agree well with theoretical predictions. Our scheme is compatible with cryogenic implementation by incorporating low-noise cryogenic amplifiers. While the present study focuses on the Kittel mode in a YIG sphere, the approach is extendable to systems with multiple magnon modes or alternative geometries.
	
%\vspace{.4cm}
\textit{Conclusion.}---To summarize, we realize the magnon-polariton lasing and amplification in the dissipatively coupled photon-magnon system. The dissipative cooperativity is found to carry clear physical significance in characterizing the system state. In particular, the condition $C_\Gamma = 1$ precisely defines the lasing threshold, which also coincides with the formation of a perfect FW-BIC. Dissipative coupling offers distinct advantages in enabling single-mode operation and tunable control of system dynamics. The lasing threshold can be further reduced by increasing the external dissipation of both modes. Our work paves the way for further exploration of non-Hermitian and nonlinear physics via dissipative coupling, including gain-induced phase transitions~\cite{Yanglan-14,Liu-2021,EI-18}, nonlinear gain saturation dynamics~\cite{Yao-2023,Jinwei-23}, and synchronization phenomena driven by non-Hermitian effects~\cite{qi-2024,Zhirov-08,Moreno-24}.

This work is supported by the National Key Research and Development Program of China (No.~2022YFA1405200 and No.~2023YFA1406703), the National Natural Science Foundation of China (No.~$92265202$, No.~$12174329$, and No.~$123\rm{B}2064$), and the Science and Technology Project of Zhejiang Province (No.~$2025\rm{C}01028$).

\textit{Data availability}---The data that support the findings of this Letter are openly available \cite{Dataset}.	
\appendix


\begin{thebibliography}{56}
\bibitem{note1}
The resonance frequency of the magnon mode in ferrites can be tuned from tens of megahertz to tens of gigahertz using a standard laboratory electromagnet.
\end{thebibliography}
\end{document}